\documentclass{article}
\usepackage{spconf,amsmath,graphicx}
\usepackage{algpseudocode}
\usepackage{xcolor}
\usepackage{textcomp}
\usepackage{algorithm}

\usepackage{booktabs}
\usepackage{url}

\title{Meta-Learning Approaches for Improving Detection of Unseen Speech Deepfakes}

\name{Ivan Kukanov$^{1}$, Janne Laakkonen$^{2}$, Tomi Kinnunen$^{2}$, Ville Hautam{\"a}ki$^{2}$}
\address{\fontsize{11pt}{10pt}\selectfont $^{1}$ KLASS Engineering and Solutions, Singapore \
\fontsize{11pt}{10pt}\selectfont $^{2}$ University of Eastern Finland, Finland}

\begin{document}

\ninept
\maketitle
\begin{abstract}
Current speech deepfake detection approaches perform satisfactorily against known adversaries; however, generalization to unseen attacks remains an open challenge. The proliferation of speech deepfakes on social media underscores the need for systems that can generalize to unseen attacks not observed during training. We address this problem from the perspective of meta-learning, aiming to learn attack-invariant features to adapt to unseen attacks with very few samples available. This approach is promising since generating of a high-scale training dataset is often expensive or infeasible. Our experiments demonstrated an improvement in the Equal Error Rate (EER) from 21.67\% to 10.42\% on the InTheWild dataset, using just 96 samples from the unseen dataset. Continuous few-shot adaptation ensures that the system remains up-to-date.
\end{abstract}
\section{Introduction}


\emph{Speech deepfakes} are artificial or manipulated (fake) speech samples generated using deep learning techniques. Although the underlying methodologies, {\em voice conversion} (VC) and \emph{text-to-speech} (TTS), serve the needs for many useful applications ranging from smart assistants and audiobooks to public announcement systems and games, the term 'deepfake' itself tends to be associated with exploitative use cases, including spoofing voice biometric systems \cite{Wu2015-survey} and telephony fraud \cite{Arizona2023-kidnapping,cnn2024-hongkong-deepfake}.

By taking a source speaker's utterance as an input, VC aims at converting the utterance to sound like the target speaker, but with the original content. TTS, in turn, generates speech from a text input. Concerning \emph{voice} characteristics, the early TTS systems were developed for a small and closed population of `stock speakers`. Recent advances in neural waveform models~\cite{Oord2016wavenet} and deep speaker embedding~\cite{desplanques2020-ECAPA} have enabled conditional speech generation by using target speaker embeddings as an additional input. This has greatly expanded the flexibility of adapting TTS, in principle, to anyone's voice using limited training data from that person. These \emph{few-shot voice cloning} approaches, such as VALL-E~\cite{wang2023-neural-codec} and AudioBox~\cite{vyas2023audiobox}, produce high-quality artificial speech in the targeted person's voice.

\begin{figure}[!htb]
\vspace{-0.1 in}
\centerline{\includegraphics[scale=0.55]{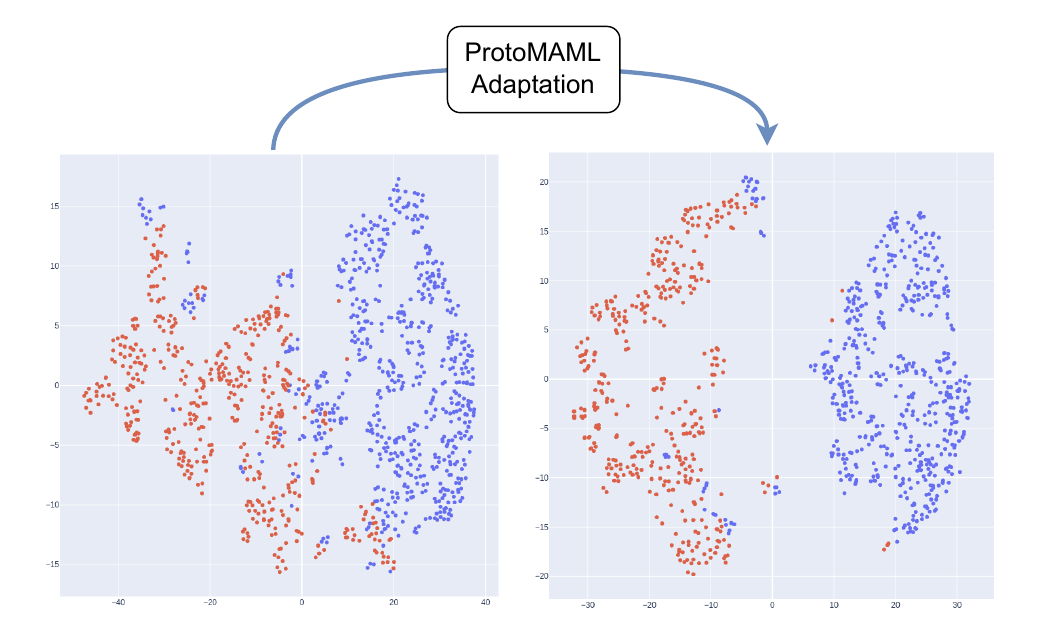}}
\caption{T-SNE visualization of model embeddings, before and after few-shot ProtoMAML adaptation on FakeAVCeleb~\cite{Hasam2021FakeAVCeleb} corpus.}
\label{fig:protomaml_adaptation}
\end{figure}

With the imminent threat posed by deepfakes, especially in telephony, teleconferencing, and social media domains, there is a timely need to develop solutions for \emph{detecting} speech deepfakes reliably. Despite the recent, popularized `deepfake' terminology, this is \emph{not} a new task for the speech community. On the contrary, speech deepfakes have received considerable attention throughout the past decade in the context of \emph{spoofing} voice biometrics (automatic speaker verification or ASV) using artificial or modified inputs. The attack vectors of biometric systems have been known for a long time~\cite{Ratha2001-enhancing}.  For speech, analysis of vulnerabilities and development of deepfake detectors have been spearheaded by the ASVspoof initiative~\cite{wu2015_ASVspoof}, a challenge series that provides common evaluation data, protocols, and performance metrics for comparing alternative speech deepfake detectors. In this context, a classifier aimed at detecting speech deepfakes is known as a (spoofing) countermeasure (CM). 
The two subsystems are assumed to produce (class-conditionally) independent decisions, leading to a cascaded spoofing-robust classifier architecture. An associated performance metric \cite{Kinnunen2020-tandem-fundamentals} measures the detection cost of the combined system. To sum up, deepfake detector is a spoofing countermeasure. While originating from different contexts, they are the same task -- differentiating spoofed (deepfake) from real (bonafide) speech utterances.  



The {\em general goal} of machine learning is that learned model parameters work well for unseen data~\cite{Finn2017modelagnostic}. By "working well" we mean that the performance on the unseen data is similar to the validation set. The main assumption behind the {\em generalization goal} is that essentially all samples in the training set, validation set, and the unseen data are all independent and identically distributed ({\em iid}) samples from the same original dataset.  In the case where we cannot guarantee that unseen samples are samples from the same distribution, we talk about the {\em domain generalization}  task. In the first glance, the task appears not to be possible, how one can train a model for data it cannot observe? A slightly easier variant of the domain generalization problem is the so-called {\em few-shot learning} task~\cite{Triantafillou2020Meta}, where we allow few samples to be observed from the unknown dataset, such as example audio files of the new speech deepfake attack type. According to \cite{Wang2021GeneralizingTU} three broad categories of approaches are currently being investigated for the domain generalization: 1) {\em data manipulation}, such as different data augmentations, 2) tools from {\em representation learning}, such as disentanglement learning and~3) {\em learning based methods}.  
Data manipulation methods, such as data augmentation, adversarial data augmentation and different data generation methods aim to increase the training dataset size and at the same time hope that generated samples match the unseen data. Representation learning, on the other hand, aims to design a model where the latent space is disentangled leading to better generalization performance~\cite{montero2021the}.

{
{




Learning-based methods~\cite{Finn2017modelagnostic, Triantafillou2020Meta}, on the other hand, approach the domain generalization task from the point of view of explicitly learning a model that generalizes well. We start with an assumption that data cannot completely change from the validation set to the unseen data, there has to be some {\em invariant} that once found can be utilized to generalize to the unseen data. For example, we assume that even in the unseen data there will be bonafide samples and also samples from some, possibly unknown fake attacks. The simplest learning-based method is to use domain adversarial training~\cite{Ganin2016}, where one trains two classifier heads for the same backbone, one for the bonafide/deepfake classifier, and the other that will classify attack types. The learning goal is to minimize loss on the bonafide/deepfake head while maximizing loss on the attack-type head~\cite{Xie2024}. The goal of the learning procedure is to force the backbone to be attack-type agnostic, at least to the known attack types.

In this work, we utilize meta-learning~\cite{Finn2017modelagnostic} based domain generalization schemes. In meta-learning the aim is to explicitly minimize the generalization loss by the attack type for testing and updating the parameters based on the loss on the few-shots of the unseen cases. The final result is, hopefully, a model that generalizes well to an unknown attack type. We are not the first to use meta-learning in speech deepfake detection task. In~\cite{Hansen2022}, the authors used meta-learning, but not in the domain generalization context.  We notice that observing just a few examples, less than a hundred, from the unseen attack type can significantly improve speech deepfake detection performance. 
To summarize questions we study in this work are:
\begin{itemize}
    \item Two meta-learning approaches are explored to address the generalization challenge by adapting to unseen fake speech data with limited samples setup.
    \item Explored a minimal number of samples needed for observing a trend in performance improvements as in-domain as well as out of domain datasets.
    \item Explored the cross-corpus generalization improvements, which makes meta-learning as an option for domain adaptation with limited training samples from unseen fake attacks with no access to TTS/VC generator.
\end{itemize}


\section{Related Work}
Most of the speech deepfake research has focused on training and testing on the same corpus. While sometimes test attacks might not be the same as in the training portion, the conditions are similar enough that reported EERs are now less than 1\%. Cross-corpus performance was systematically evaluated in studies such as~\cite{muller2022does, Shim2023} with the results that countermeasures do not generalize well to unseen corpora. 
Regularization is one of the approaches that have been tried to solve cross-corpus generalization problem, such as in~\cite{Chen2020} where authors used large margin cosine loss and frequency masking augmentation. Another approach was taken in~\cite{Ma2021}, where authors finetuned model with unseen data but used continual learning to ensure that the model still worked on the original corpus. The idea is that parameters should not diverge too far from the base model and thus still work in the original corpus. Main conceptual difference to the proposed is that our aim is to use minimal amount of unseen data in the adaptation. 
In~\cite{Kawa2022}, the authors approached generalization by combining three corpora and performing 5-fold cross-validation and using double frontend LCNN model. One approach to improving generalization is to train a massively large SSL model and then finetune it with some speech deepfake corpus, with the idea that such a model will then generalize to unseen corpus~\cite{oneata2023generalisable, Wang2023}. In the present work we take a large SSL model as a basis, but instead of finetuning the whole network we finetune only a small part using meta-learning based strategy.

\section{Meta-learning}
The basic idea of meta-learning~\cite{Finn2017modelagnostic} is that given multiple different tasks, it is beneficial to learn a general set of parameters that are shared by all the tasks. After that, learning (adapting) an individual task becomes easier. The whole dataset is then divided into multiple sets, such as $\mathcal{D} = \{ \mathcal{T}_t\}_{t=1}^T$ is the training set for the tasks of interest and $\mathbf{x}_i$ is the input data and $y_i$ is the corresponding label. Each task $\mathcal{T}_t$ consists of \textit{support} $\mathcal{S} = \{S_i\}_{i=1}^N$ and \textit{query} $\mathcal{Q} = \{Q_i\}_{i=1}^N$ sets with $N$-classes, where $S_i = \{(\mathbf{x}_j, y_j)\}_{j=1}^K$ is $K$-shot samples from class $i$.

The meta-learning process comprises two phases: \textit{meta-training} and \textit{meta-testing}. In \textbf{\textit{meta-training}}, the model encounters multiple tasks $\mathcal{T}_t$, each divided into a support set $\mathcal{S}_t$ and a query set $\mathcal{Q}_t$. The model adapts to each task using the support set through a few gradient updates; then it is evaluated on the query set. The goal is to optimize the model's initial parameters on the support sets to perform well on the corresponding query sets. During \textbf{\textit{meta-testing}}, the model's generalization ability is assessed on new, unseen tasks, similarly divided into support and query sets. This time, the model adapts to these new tasks using few-shots from the support set; and it is evaluated on the query set representing the whole evaluation set. The tasks alternation, during meta-training, learns the ability to adapt to new tasks quickly.

\section{Domain Adaptation In Speech Deepfakes}


Meta-learning, or "learning to learn," has emerged as a powerful paradigm for enabling models to learn new tasks quickly with minimal data~\cite{Finn2017modelagnostic}. Meta-learning is valuable in fields where data is scarce or expensive to obtain. Among the plethora of meta-learning algorithms, two have shown notable promise in current research: {\em prototypical networks} (ProtoNet) \cite{Snell2017Prototypical}, {\em model-agnostic meta-learning} (ProtoMAML) \cite{Triantafillou2020Meta}. This section provides an overview of these methods, highlighting their foundational concepts.

\subsection{Prototypical network} 

\begin{figure}[!htb]
\vspace{-0.1 in}
\centerline{\includegraphics[scale=0.55]{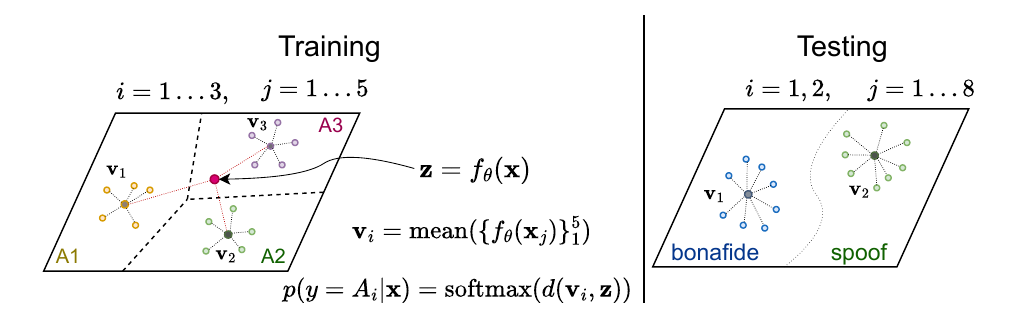}}
\caption{ProtoNet intuition: during training the optimal centroinds are learned for every subset of tasks, the adaptation to a new tasks is done with few-shots to compute prototypical vectors of new tasks. For testing, the distance to the nearest centroid is computed to known  class.}
\label{fig:protonet_intuition}
\end{figure}

Prototypical network (ProtoNet)~\cite{Snell2017Prototypical} offers an effective approach to few-shot learning tasks. It operates in the principle of \textit{learning a metric space} in which classification can be performed by computing distances to prototype representations of each class. This method excels in scenarios where the goal is to rapidly adapt to new tasks with a limited number of examples~\cite{Snell2017Prototypical}. 

Prototypical network computes a prototype centroid vector $\mathbf{v}_i$ for every class $i$ based on samples $\mathbf{x}_j$ from a support set $S_i$:
\begin{equation}
\mathbf{v}_i = \frac{1}{\big{|}S_i\big{|}} \sum_{\mathbf{x}_j \in S_i} f_{\theta}(\mathbf{x}_j),
\end{equation}
where $N$-way and $K$-shot indexes are $i = \overline{1, N}$ (class index) and $j = \overline{1, K}$ (sample index). Each query example $\mathbf{x}^*$ is assigned to the class whose prototype is the nearest to it, using squared Euclidean distance for measurement. Specifically, the probability of predicting class $i$ for the input sample $\mathbf{x}^*$ is determined
\begin{equation} 
\label{eq:softmax_protonet}
p(\mathbf{y}^* = i |\mathbf{x}^*) = \text{softmax}\bigl(-||f_{\theta}(\mathbf{x}^*) - \mathbf{v}_i ||^2_2 \bigr),
\end{equation}
where the squared Euclidean distance is between a sample embedding and a prototype vector.

\subsection{ProtoMAML} 

The hybrid approach ProtoMAML integrates the prototypical concept into the MAML structure to improve the adaptation step in a few shots \cite{Triantafillou2020Meta}. Compared to non-parametric ProtoNet, ProtoMAML is optimization-based, i.e. updates the network parameters during the adaptation step to a new task. ProtoMAML aims to optimize a model's parameters such that a small number of gradient updates will lead to significant improvement on a new task.

\begin{algorithm}
\caption{ProtoMAML Algorithm}
\label{alg:protomaml}
\begin{algorithmic}[1]
\State \textbf{Input:} meta-training set $\mathcal{D}_{\text{tr}}$, inner rate $\alpha$, outer rate $\beta$
\State Initialize model parameters $\theta$
\While{not converged}  \textit{(outer loop)}
    \For{each task $\mathcal{T}_t \sim \mathcal{D}_{\text{tr}}$} \textit{(inner loop)}
        \State Sample support $\mathcal{S}$ and query $\mathcal{Q}$ from $\mathcal{T}_t$
        \State Compute prototypes $\mathbf{v}_i$ for each class $i$ in $\mathcal{S}$ using $\theta$
        \State Evaluate $\nabla_{\theta} \mathcal{L}_{\mathcal{T}_t}(\mathcal{S}; \theta)$ using $\mathcal{S}$ and $\mathbf{v}_i$
        \State Compute adapted parameters with gradient descent: 
            $\centerline{$\theta'_t = \theta - \alpha \nabla_{\theta} \mathcal{L}_{\mathcal{T}_t}(\mathcal{S}; \theta)$}$
        \State Compute loss on $\mathcal{Q}$ using $\theta'_t$ and $\mathbf{v}_i$: $\mathcal{L}_{\mathcal{T}_t}(\mathcal{Q}; \theta'_t)$
    \EndFor
    \State Update $\theta \gets \theta - \beta \nabla_{\theta} \sum_{\mathcal{T}_t} \mathcal{L}_{\mathcal{T}_t}(\mathcal{Q}; \theta'_t)$
\EndWhile
\end{algorithmic}
\end{algorithm}

ProtoMAML networks are interpreted as ProtoNet with a linear classifier applied to a learned representations $f_{\theta}(\mathbf{x})$. In ProtoMAML~\cite{Triantafillou2020Meta}, the squared Euclidean distance from (\ref{eq:softmax_protonet})
\begin{equation}
   -||f_{\theta}(\mathbf{x}) - \mathbf{v}_i ||^2_2 = 2\mathbf{v}_i^Tf_{\theta}(\mathbf{x}) - ||\mathbf{v}_i||^2 + \text{const},
\end{equation}
where we denote $\mathbf{W}_{i,.} = 2\mathbf{v}_i$ and $b_i = -||\mathbf{v}_i||^2$, which gives trainable output linear layer $\mathbf{W}f_{\theta}(\mathbf{x}) + b$.



\section{Experimental setup}

\subsection{Dataset}


{\bf Training Corpus:} The ASVspoof2019 LA~\cite{ASVspoof2019} dataset is used to train all systems in this study, it has 6 attacks in training/validation and 13 attacks in evaluation split. Six attacks are designated as known spoofing systems (A01-A06), include two VC and four TTS systems. In the evaluation set, A07-A19 (except A16 and A19) are the eleven unknown spoofing systems, and A16 and A19 are the known reference systems using the same algorithms as A04 and A06. We use spoofing attack labels for training multi-class classification with meta-learning methods. Though, the adaptation and the evaluation is done in a binary classification way (\textit{bonafide} \& \textit{spoof}) for all systems.


{\bf Testing Corpora:} We evaluate systems' performance on multiple test sets to measure generalization in both seen and unseen domains. It comprises conventional evaluation splits of ASVspoof2019 LA, ASVspoof2021 LA and DF tasks~\cite{ASVspoof2021}. Two out-of-domain test sets were In-The-Wild~\cite{muller2022does} and FakeAVCeleb~\cite{Hasam2021FakeAVCeleb} corpora, both of whose domains differ from that of ASVspoofs. In particular, In-The-Wild contains bonafide and spoofed utterances from 54 English-speaking celebrities, collected from the Internet. FakeAVCeleb is based on VoxCeleb2~\cite{Chung2018VoxCeleb2} dataset and has five English accents; all deepfake utterances are generated with transfer learning-based real-time voice cloning~\cite{Jia2018Transfer}. 

\subsection{Training Strategy}
We explore the top performing fake speech detection model derived from ASVspoof community \cite{Tak2022Automatic,Wang2023Can}. It is based on SSL \textit{front-end} and graph neural network \textit{back-end}.

\textbf{Self-supervised learning (SSL) model.} The system Wav2Vec-AASIST in Table~\ref{tab:results} is trained using Wav2Vec 2.0 XLSR-53 with 1024 output embedding dimension coupled with an integrated spectro-temporal graph attention
network (AASIST)~\cite{Tak2022Automatic} as a back-end. In the Wav2Vec-AASIST* scenario, the SSL model is frozen during training, only parameters of the graph attention are trained. We train the Wav2Vec-AASIST and the Wav2Vec-AASIST*  model using the same setup. We use Adam~\cite{Kingma2015Adam} optimizer with an initial learning rate of $1 \times 10^{-6}$. Negative log-likelihood loss is optimized with 2 class \textit{logSoftmax} outputs for \textit{bonafide} and \textit{spoof}. Training stops using early stopping if there is no improvement in validation performance for more than 15 epochs. The best performing model on the validation set is used for testing.

\textbf{Meta-Learning: ProtoNet and ProtoMAML.} In this experiment we focus on few-shot classification task, where the classes in the training and test sets are different (or partially unseen) as in the case of ASVspoof19 dataset which has different type of attacks in training and evaluation splits. 
In meta-training phase, we randomly sample 3 classes out of 7 classes (A01-A06 and bonafide) for each task. However, at meta-test time, we classify between \textit{spoof} and \textit{bonafide}.

The Wav2Vec-AASIST* network backbone was used in meta-learning. The Wav2Vec 2.0 checkpoint is used to initialize front-end weights and it is kept frozen for all meta-learning experiments. The graph attention network back-end (AASIST) is randomly initialized with values using a Xavier normal distribution~\cite{Glorot2010Understanding}. Only AASIST parameters are optimized during training. The training was done in few-shot manner with 3-class and 5-shot setting. 
That is, every iteration task sampler randomly picks 5 samples of 3-classes for support and query sets from ASVspoof2019 training set. For each update step, we train on support batch and validate on query batch. Typically, it is advised to keep the number of shots during training equal to the number intended for adaptation/evaluation~\cite{Snell2017Prototypical}. However, we explore the number of shots as a hyper parameter later in the adaptation section to obtain the best-performing model.

The \textbf{ProtoNet} learning strategy was applied to Wav2Vec-AASIST* which outputs $64$-dimension embedding for calculating prototype vector for every class. On each support batch, we calculate an averaged prototype vector for each class, then we calculate the \textit{logSoftmax} of the squared Euclidean distance between the prototype vector and each sample in a query batch. The \textit{cross entropy} loss is optimized to predict the most probable sample class which embedding is the closes to the corresponding prototype. The loss is minimized with AdamW~\cite{Loshchilov2019Decoupled} optimizer with the initial learning rate of $1\times10^{-3}$. The learning rate was scheduled with cyclic policy (CyclicLR)~\cite{Leslie2017Cyclical}, specifically triangular mode, which cycled in a range between $[1\times10^{-6}, 1\times10^{-3}]$, and the triangular step size is $8$ epoch. The training is carried out for $200$ epochs and the best-performing model on the validation set is selected for evaluation.

The \textbf{ProtoMAML} training has inner and outer optimization loops in Algorithm \ref{alg:protomaml}. We use the same architecture Wav2Vec-AASIST* and feature space size of $64$ as for ProtoNet, the outer loop gradients are accumulated over 4 batches. The optimizer and the learning rate scheduler for the outer optimization are the same as for ProtoNet.
The inner optimization is performed with SGD and update learning rate is set to $0.1$, which is much higher than the outer loop learning rate of AdamW optimizer. It was suggested in~\cite{Triantafillou2020Meta} that more aggressive inner-loop learning rate improves MAML’s performance significantly. Another crucial hyperparameter is the number of inner loop updates, it depends on the similarity of our training tasks. In the training, we notice that a single inner loop update achieves the best performance and learns faster. 


\begin{table*}[!htb]
\footnotesize
\centering
\caption{Comparison of binary classification vs ProtoNet and ProtoMAML adaptation with 96 and 256 samples. Shots per class column denotes the number of samples used to update the model. Performance is reported in terms of EER (\%).}
\setlength{\tabcolsep}{1.7mm}{
\begin{tabular}{lccccccc}
\hline
\toprule
Model & \multicolumn{1}{c}{\shortstack{trainable\\ params.}} & \multicolumn{1}{c}{\shortstack{shots\\ per class}} & \multicolumn{1}{c}{ASV19:LA} & \multicolumn{1}{c}{ASV21:LA} & \multicolumn{1}{c}{ASV21:DF} & \multicolumn{1}{c}{InTheWild} & \multicolumn{1}{c}{FakeAVCeleb} \\ \hline
\midrule
Wav2Vec-AASIST~\cite{Tak2022Automatic} & 317,837,834 & - & 0.84 & 7.04 & 4.49 & 13.93 & 7.44\\ \hline
Wav2Vec-AASIST* & 447,242 & - & 1.46 & 11.39 & 8.35 & 21.67 & 6.07\\
ProtoNet\_256	& 457,224 & 256 & 1.68 &	\textbf{4.55} &	7.65 & 20.94& 10.29 \\
ProtoNet\_96	& 457,224 & 96 & 1.68 & 4.56 & 7.65 & 21.04 & 10.35 \\
ProtoMAML\_96	& 457,224 & 96 & \textbf{1.35} & 6.39 & \textbf{6.05} & \textbf{16.29} & \textbf{5.55} \\
\bottomrule
\hline
\end{tabular}
}
\label{tab:results}
\end{table*}

\subsection{Few-shot Adaptation and Evaluation}

To evaluate the adaptation capability, we adapt \mbox{ProtoNet} and \mbox{ProtoMAML} to new deepfakes from unseen domain. We randomly select $k$ speech samples from spoof and bonafide classes to extract the prototypes and assess the model's equal error rate (EER) performance metric using the remaining samples of the evaluation set. Meta-testing process is similar to a meta-training, the support set is used for adaptation on unseen domain and the remaining of the evaluation dataset is a query set. 

We repeat random sampling of support set for 9 times. Each time we run adaptation followed by an evaluation on each evaluation set. We calculate mean and standard deviation confidence boundaries, see Figure \ref{fig:protonet_results} and \ref{fig:protomaml_results}. The averaged performance across all support sets indicates the expected efficacy of ProtoNet and ProtoMAML approaches when exposed to a limited number of examples per class. During training, we used $k=5$ shots, however, for the testing phase, we vary the number of shots to examine its impact on performance. We experiment with $k$ varying from $2-256$ for ProtoNet and from $2-96$ for ProtoMAML. Computational restrictions allowed us to explore only up to this number of shots.

The ProtoMAML is evaluated the same way as ProtoNet, specifically by selecting 2-class and $k$-shot random examples from the test set to form a support set, while the remainder of the evaluation dataset serves as the query set. However, unlike simple calculating prototypes for all classes in support set, ProtoMAML requires fine-tuning a separate model for each support set inside the inner loop in Algorithm \ref{alg:protomaml}. This adaptation step makes the evaluation process resource-intensive compared to ProtoNet, however, more adaptable to the new acoustic environment. In contrast to training phase, it is recommended to use many more inner-loop updates during adaptation on the support set. To obtain results in Figure \ref{fig:protomaml_results} and Table \ref{tab:results}, we used $25$ inner-loop adaptation steps for all ProtoMAML models. Finally, in Figure \ref{fig:protomaml_nsteps}, we investigate the effect of number of adaptation steps on performance.

\section{Results}

\begin{figure}[!htb]
\vspace{-0.1 in}
\centerline{\hspace{-0.2cm}\includegraphics[scale=0.4]{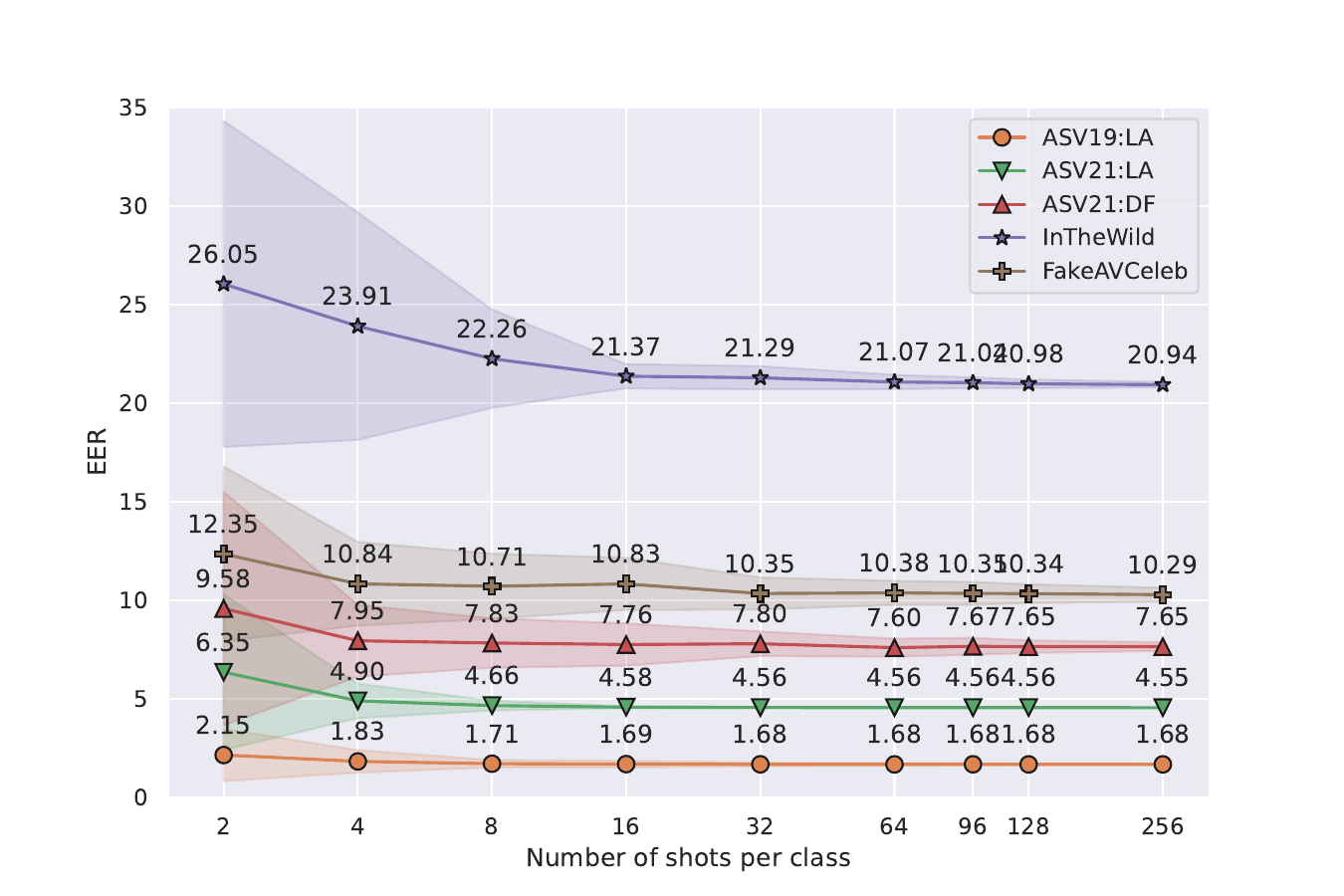}}
\caption{ Explore ProtoNet few-shot adaptation with 2-256 shots per class. Horizontal axis is log-scaled.}
\label{fig:protonet_results}
\end{figure}

\begin{figure}[!htb]
\vspace{-0.1 in}
\centerline{\hspace{-0.2cm}\includegraphics[scale=0.4]{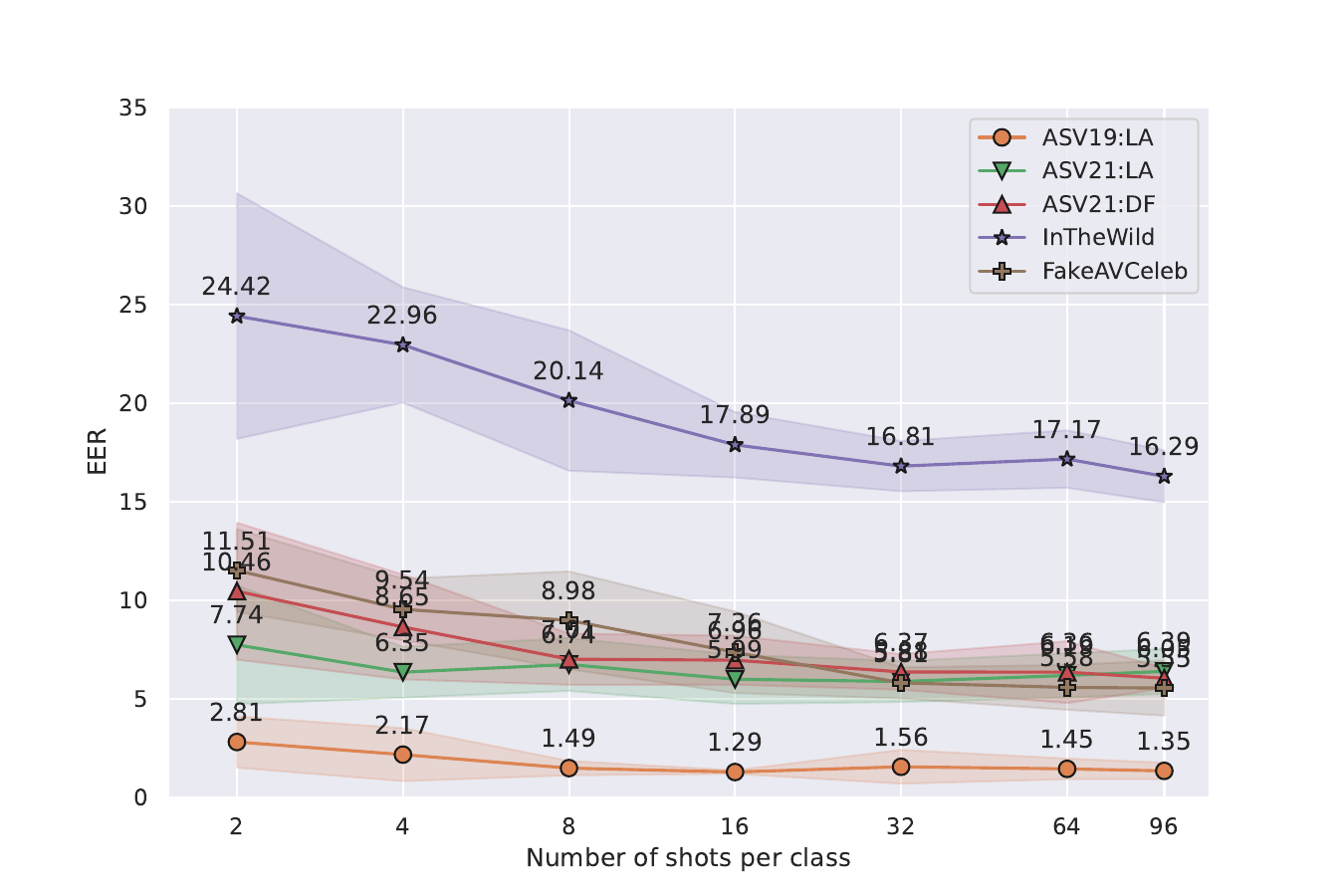}}
\caption{ Explore  ProtoMAML few-shot adaptation with 2-96 shots per class. Horizontal axis is log-scaled.}
\label{fig:protomaml_results}
\end{figure}

\textbf{Number of adaptation shots.} Number of few-shot adaptation for across datasets is shown in Figures \ref{fig:protonet_results} and \ref{fig:protomaml_results}. It shows that increasing the number of examples in a support set for adaptation on unseen domain improves the performance and facilitate the investigation at scales. The $k$-shot from $2-256$ is explored for ProtoNet; and from $2-96$ is for ProtoMAML. 
For both ProtoNet and ProtoMAML, the EER decreases the more adaptation samples are introduced. We see dramatic exponential error drop after introducing up to 16-shot per class, e.g., for ProtoNet, the EER on InTheWild dropped from $26.05\%$ to $21.37\%$ and for ProtoMAML from $24.42\%$ to $17.89\%$. After 16-shot the improvement stagnates across all evaluation corpus, and the variation drops between different support samples, the confidence boundaries narrows down.
The most challenging dataset for adaptation is InTheWild, especially adapting on small support set below 16-shot. However, even with just 256-shot per class, ProtoNet brings EER to $20.94\%$ and ProtoMAML brings to $16.29\%$ on 96-shot. 

From Table \ref{tab:results}, due to few-shot adaptation, meta-learning approaches outperform supervised approach Wav2Vec-AASIST* with almost the same number of trainable parameters across all datasets. The ProtoMAML, as optimization-based adaptation, outperforms non-parametric ProtoNet on majority of datasets, except ASV21:LA.

\textbf{Number of adaptation steps.} Analysing the number of adaptation steps for ProtoMAML\_96, Figure \ref{fig:protomaml_nsteps}, we observed exponential performance improvement on InTheWild dataset. Finetuning ProtoMAML on 96-shot and 2-class (\textit{bonafide} and \textit{spoof}) support set, the EER dropped significantly from $17.61\%$ to $10.42\%$ when using 5K adaptation steps. We experimented with the adaptation steps only on InTheWild datasets as the most challenging for adaptation, other datasets require further investigation.

In summary, it is evident that ProtoMAML standing out over ProtoNet when it comes to adaptation step tuning. However, a notable drawback of ProtoMAML is its significant computational requirements. ProtoNet, with its simplicity, offers a robust baseline for ProtoMAML and may be the preferred choice in scenarios where resources are constrained.



\begin{figure}[!htb]
\vspace{-0.1 in}
\centerline{\hspace{-0.2cm}\includegraphics[scale=0.4]{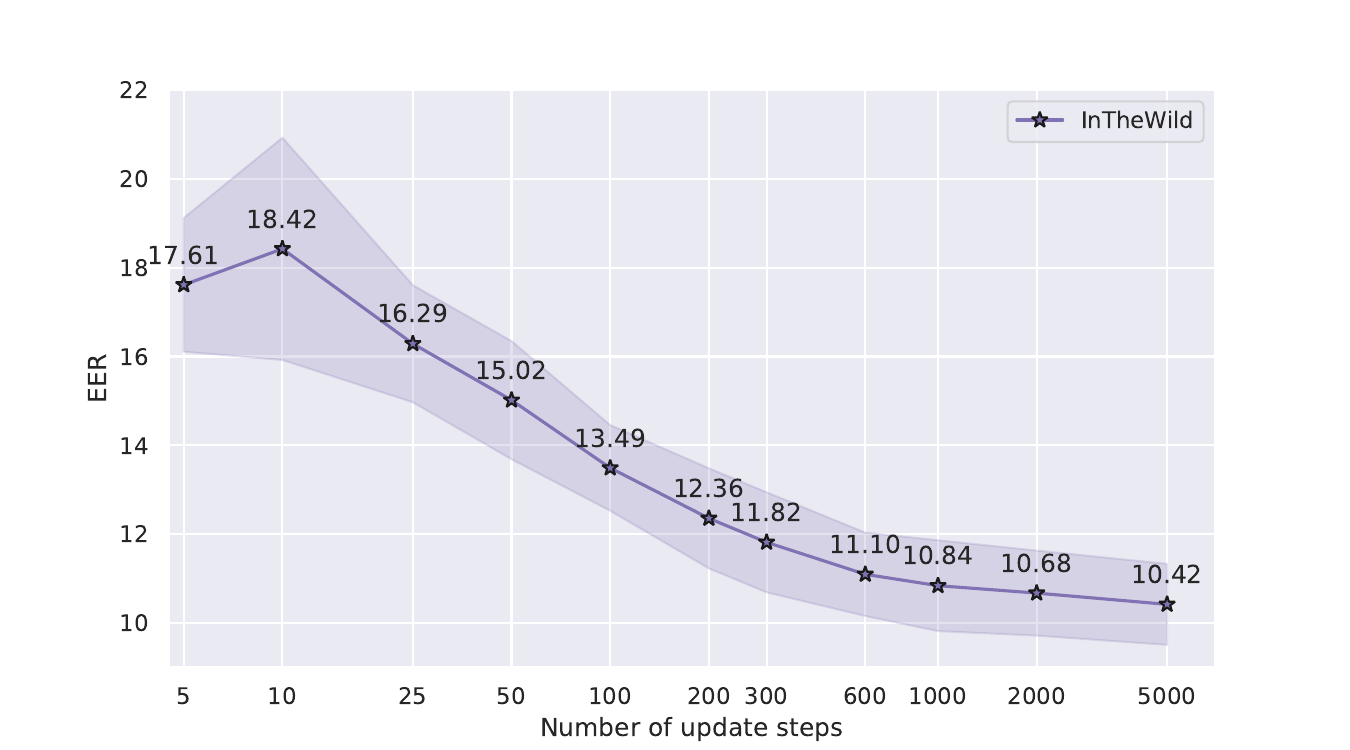}}
\caption{ Explore  ProtoMAML\_96 adaptation steps on 96-shots per class. Horizontal axis is log-scaled.}
\label{fig:protomaml_nsteps}
\end{figure}

\section{Conclusions}
Previous work had already shown that in the cross-corpus task, where a speech deepfake detector is trained using one corpus and applied to another corpus, the generalization performance degrades considerably. We confirm it by showing that the performance of the Wav2Vec-AASIST decreases from $0.84\%$ EER to $13,93\%$ EER when applied to the InTheWild corpus. Our aim to solve the generalization problem is to use meta-learning in the way of adapting the trained model to a new corpus. Specifically, we used ProtoNet and ProtoMAML training and adaptation schemes. Resulting in that when we use only 96 samples per class from the InTheWild corpus, we are able to push the EER to $10.42\%$. This result was obtained with $0.14\%$ of the trainable parameters used in the baseline model. 

This is a promising direction for continuous system adaptation when we have few open-sourced samples from a new realistic synthetic speech attacks. In that case, generating a high-scale training dataset is expensive or not feasible, however, running few-shot adaptation will, at least,  guarantee that the system is up to date. The future goal of our work is to achieve a reasonable level of performance without sampling from the evaluation set.



\section{ACKNOWLEDGMENTS}
The work has been partially supported by the Academy of Finland (Decision \textnumero 349605, project "SPEECHFAKES"). The authors wish to acknowledge CSC – IT Center for Science, Finland, for computational resources.



\bibliographystyle{IEEEbib}
\bibliography{strings,refs}

\end{document}